**Charge-state control of carbon-related optical absorption in AlN**

Helen C. Robinson[1], Daniil Danilin[1], Md Shafiqul Islam Mollik[1], Darshana Wickramaratne[2], John L Lyons[2], Vladimir Fedorov[1], Sergey Mirov[1], and M E Zvanut[1]

[1]University of Alabama at Birmingham, Department of Physics, Birmingham AL 19233 USA

[2]Center for Computational Materials Science, US Naval Research Laboratory, Washington, D.C. 20375 USA

Sub-bandgap optical absorption in AlN between 2 eV and 4 eV is widely observed, but its microscopic origin remains contested. Using photo-induced electron paramagnetic resonance (photo-EPR) and optical absorption spectroscopy on the same samples, we demonstrate a correlation between this absorption band and the neutral charge state of substitutional carbon on the nitrogen site ($C_N$). Hybrid functional calculations of the optical absorption spectra show that a transition involving $C_N$ and the valence band occurs near 3.3 eV, which agrees well with a peak identified within the measured optical absorption between 2 eV and 4 eV. This conclusion requires the combined ability to manipulate the charge state of carbon using photo-EPR and to use first-principles calculations of the absorption line shape that account for the dispersion of the valence band and the energy dependence of the optical matrix elements.

A 6.2 eV ultrawide bandgap makes AlN an attractive candidate for ultraviolet photonics and high-power electronics [1-5]. However, AlN-based devices are susceptible to incorporation of unintentional impurities, such as carbon, oxygen, and silicon, as well as intrinsic defects, all of which could introduce sub-bandgap optical absorption that can limit UV transparency and device performance. While optical absorption (OA) spectroscopy is routinely used to evaluate AlN substrates, linking absorption features to specific defects has proven remarkably challenging [6-8]. Apart from a band near 4.7 eV convincingly attributed to carbon on the nitrogen site ($C_N$) [9,10,11], the origins of other below-bandgap absorption in AlN remains poorly understood. Given the well-documented impact of carbon on the electrical [1-3], thermal [12], and optical [10,13,14] properties of AlN, a better understanding of the effects of carbon on the optical absorption is needed.

Bulk AlN substrates routinely exhibit a broad absorption extending from 2 eV to 4 eV containing several overlapping peaks with positions and relative intensities which vary among samples and illumination conditions [8,11, 14, 15]. For example, absorption near 2.8 eV has been attributed by some to nitrogen vacancies ($V_N$) and others to aluminum vacancies ($V_{\mathrm{Al}}$) [6-8, 13,14]. A peak near 3.4 eV, which has seldom been experimentally resolved, has been assigned to isolated $V_{\mathrm{Al}}$ or $V_{\mathrm{Al}}$-related complexes [8, 14-17]. Attempts to decompose the broad absorption band into individual Gaussians yielded a peak at 2.78 eV which was assigned to a transition between an $V_{Al}$ and $V_N$ and one at 3.59 eV which was assigned to $V_{\mathrm{Al}}$ [8]. After thermal annealing, Gamov *et al.* resolved a feature near 2.6 eV and assigned it to transitions involving the $C_N$ acceptor level and the valence band [11]. The range of defect candidates that have been ascribed to the same optical absorption peak reflects the absence of independent structural or chemical identification of the responsible defects.

In this work, we show that absorption features spanning 2 eV to 4 eV are correlated with carbon and that at least one is governed by the charge state of $C_N$. Combining photo-induced electron paramagnetic resonance (photo-EPR) with OA and first-principles calculations, we demonstrate that a peak near 3.4 eV arises from transitions between the valence band and the $C_N$ acceptor level. Two non-trivial factors make this assignment possible. First, our recent EPR identification of the neutral charge state of $C_N$ in AlN [9] provides unambiguous identification of the charge state of $C_N$ prior to each measurement. Second, optical absorption measurements are combined with a pre-illumination step using light emitting diodes (LEDs) at wavelengths chosen to drive a controlled change in the charge state of $C_N$, thus enabling the absorption features to be switched on and off. Crucially we show, that by including the energy dependence of the optical matrix elements due to transitions below the valence band maximum (VBM), an absorption peak emerges which aligns well with the one seen at 3.4 eV. We attribute this peak to transitions involving $C_N$ and the valence band.

AlN substrates grown via physical vapor transport were cut into rectangular pieces measuring 0.7–1.6 cm × 0.2 cm. All samples are 500 μm thick, polished on both sides, and have a mostly uniform yellowish coloration. The concentration of Si and O donors in the various samples ranged from 2-9x$10^{18}$ $cm^{-3}$ and carbon was typically 7.5-10x$10^{19}$ $cm^{-3}$ as measured by secondary ion mass spectrometry. 10 GHz EPR measurements were performed with the magnetic field applied parallel to the c-axis of the sample and a Shimadzu UV-3101PC spectrophotometer was used to perform the optical absorption. All measurements were made at room temperature. Transmission values were converted to absorption coefficients, α, using the Lambert–Beer Law:

$$\alpha = -\frac{ln\,(T)}{z} \qquad \text{eq. 1}$$

where z is the sample thickness and T is the transmittance [18]. Transmittance was not corrected for reflection because our focus is the change in absorption rather than the absolute value of the absorption coefficient. With a sample thickness of 500 μm, the maximum measurable absorption coefficient is about 80 $cm^{-1}$.

Samples were illuminated by selected LEDs for 15 to 20 minutes for EPR and OA measurements to maximize the signal amplitudes. The output power of the LED ranged from 0.5 to 25 mW, and neutral density filters were used to ensure that the photon flux at each exposure wavelength was similar. For EPR, the LED illumination was performed in-situ by passing the light through a collimator and directing it through the slits of the EPR cavity where the entire sample is measured at each wavelength. Details of the photo-EPR measurement may be found in reference [9]. For OA, the LED exposure could not be performed in-situ and only a $mm^2$ spot on the sample is measured by the spectrophotometer. Therefore, a special sample holder was constructed to minimize the time between LED exposure and OA measurement. The holder also ensured that the spot illuminated by the LED was the same as that through which the transmittance was measured. To check for uniformity along the sample, the position of the spot was varied for both LED illumination and optical absorption measurement. It is essential to our study that the spectrophotometer light sources do not induce changes like those caused by the LED. This was demonstrated by showing that repeated OA measurements produced identical spectra. Throughout this work, LED exposure that alters the charge-state population are referred to as pre-illumination, while the subsequent spectrophotometer measurement is referred to as the OA measurement.

We perform first-principles calculations based on density functional theory (DFT) [19] using the projector-augmented wave (PAW) [20] method as implemented in the Vienna Ab-initio Simulation Package (VASP) [21]. We use the Heyd-Scuseria-Ernzerhof (HSE) hybrid functional [22] with a plane-wave energy cutoff of 500 eV. The fraction of nonlocal Hartree-Fock exchange is set to 0.33, which reproduces experimental band gaps and lattice parameters of AlN. Wavefunctions and energies for carbon substituting on the nitrogen site in AlN are taken from our previous study using the results from 288-atom supercells [9].

To resolve individual peaks in the absorption spectra, we calculate the normalized absorption line shape for transitions between the AlN valence band and the $C_N$ acceptor level as a function of photon energy $\hbar\omega$. The calculation follows the established one-dimensional (1D) configuration coordinate approach, which has been shown to give quantitative agreement with experiment when electron-phonon coupling is large, as is the case for $C_N$ in AlN [9]. One distinction from prior line shape calculations of this type is that we do not assume the optical matrix elements between the defect and band states are constant. Instead, we explicitly compute their energy dependence using HSE hybrid functional calculations on a dense 6×6×6 k-point grid, dividing the grid into groups of k-points for separate calculations and combining the results. The line shape function is

$$\alpha(\hbar\omega) \propto \sum_{i,k} \left|M_{i,k}\right|^2 \sum_{n,m} w_g \left\langle \chi_{en} \middle| \chi_{gm} \right\rangle^2 \delta(E_{ZPL} + n\Omega_e - \mathrm{m}\Omega_g + \hbar\omega). \qquad \text{eq. 2}$$

The sum over $i$ and $k$ runs over all band indices ($i$) and valence band $k$-points ($k$), $|M_{i,k}|^2$ is the momentum resolved optical matrix element squared between the unoccupied Kohn Sham state of the $C_N$ defect level (calculated in the neutral charge state) and a valence band state with a given band index and $k$-point. $\langle \chi_{en} | \chi_{gm} \rangle$ are the Franck-Condon overlap integrals evaluated at $T =$ 300 K (using a thermal occupation factor $w_g$) between vibrational levels in the ground (g) and excited (e) state where $n$ and $m$ are the phonon replicas. The Dirac delta function (which defines the energy conservation in the absorption process) contains the zero-phonon line, $E_{ZPL}$, for the absorption process which is the ionization energy of the $C_N^{-/0}$ level with respect to the VBM, $\Omega_{e,g}$ are the effective phonon energies in the excited and ground state obtained within a 1D approximation for an absorption process involving the valence band and the $C_N$ acceptor level [9], and $\hbar\omega$ is the photon energy. In our calculations the delta function is replaced by a Gaussian with a finite width.

The samples were divided into two sets: those in which the EPR spectrum due to $C_N^0$ could be seen only after illumination, referred to as set 1, and those in which the $C_N^0$ EPR spectrum is present in the dark and minimally altered by LED illumination, which we refer to as set 2. Figure 1a shows representative EPR data from a set 1 sample. The spectrum measured in the dark (black) is the same as that seen in an empty sample holder, confirming $C_N$ is EPR silent in the negative charge state. After pre-illumination with a 265 nm LED for 20 minutes (red), the characteristic spectrum of the neutral charge state of $C_N$ appears, consistent with photoionization of the occupied $C_N$ level resulting in an electron in the conduction band and conversion of $C_N$ to the neutral charge state [9]. Subsequent illumination with a 530 nm LE D quenches th e neutral $C_N$ EPR signal (blue), consistent with photoexcitation of valence band electrons that returns $C_N$ to the negative charge state. As we show below, the optical absorption spectra track this charge-state switching in direct correspondence with the EPR signal, which is the focus of the present work.

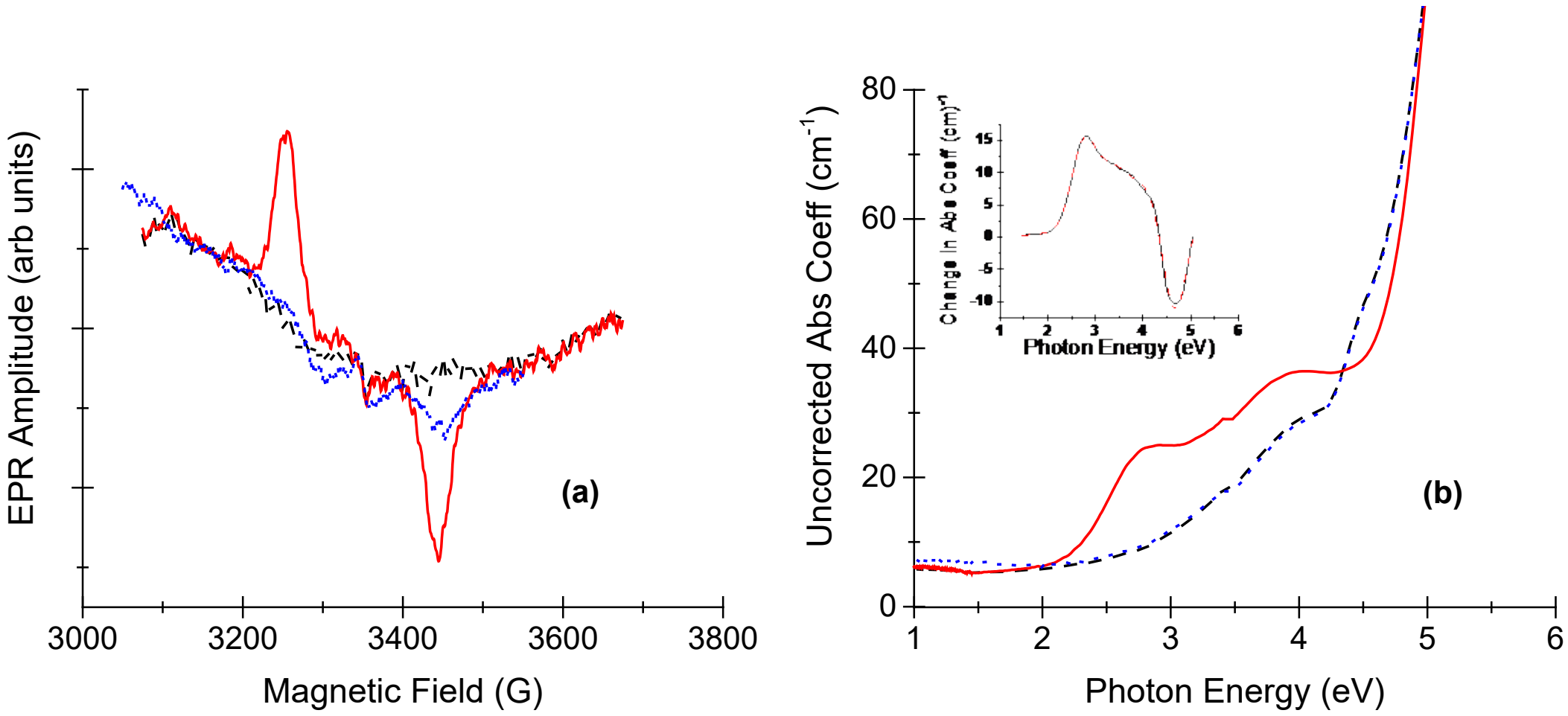

Figure 1. EPR (a) and optical absorption (b) spectra of an AlN sample from set 1: dark (black, dashed); after illumination with 265 nm LED (red, solid) and subsequent illumination with 530 nm LED (blue, dotted). INSET: Subtraction (black) of the dark OA from the 265 nm induced OA and fit (red dashed) using four gaussian line shapes.

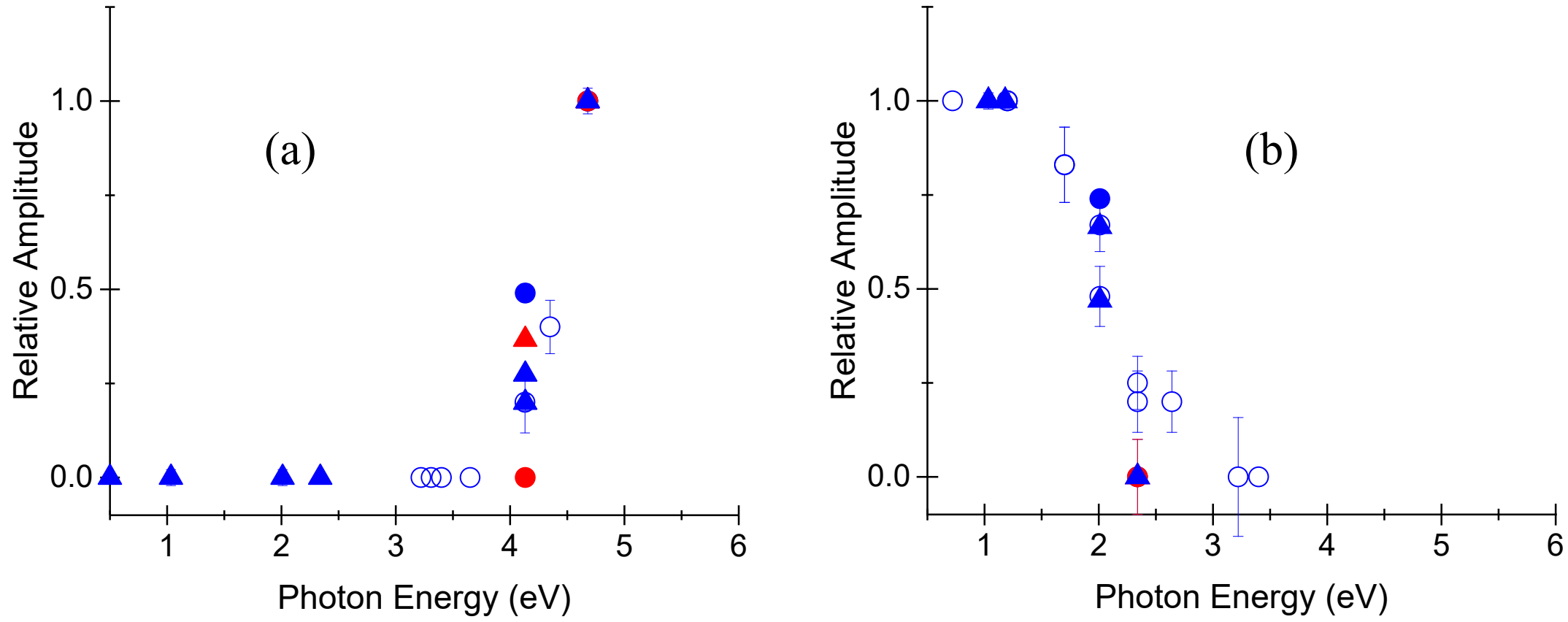

Figure 2: Relative amount of $C_N^0$ (circles) and 2-4 eV absorption coefficients (triangles) measured during pre-illumination with wavelengths from dark to 265 nm (4.7 eV) (a) and subsequent quenching of the 4.7 eV generated signal (b). The different color symbols were obtained from two different samples from set 1.

The EPR samples from set 1 were also used to measure OA using an approach similar to that described in reference [11]. Figure 1b illustrates the measured OA after the sample was in the dark for several days. The black curve illustrates the broad absorption seen in many AlN substrates, while pre-illumination at 265 nm (4.7 eV) increases the OA between 2 and 4 eV (red curve) and decreases the absorption above 4 eV, relative to the OA measured in the dark. Next, we perform pre-illumination at 530 nm (2.3 eV) and find the OA (blue curve) resembles the spectra that we initially measured when the sample was in the dark (black curve). Note that the LEDs used for pre-illumination are the same as those used to alter the $C_N$ EPR spectra described above.

To isolate the absorption induced by pre-illumination, OA spectra measured in the dark were subtracted from those obtained after LED pre-illumination, and a set of Gaussian line shapes was fit to the difference spectra. The inset of Figure 1b shows a representative spectrum after 265 nm illumination (solid black) with the fit (dashed red). Four peaks are identified: two partially resolved ones at 2.7 eV and 4.7 eV, and two unresolved peaks at 3.4 eV and 4.3 eV, with variation of less than 3% among samples. The 4.7 eV feature, which lies near the detection limit, is consistent with the absorption band attributed to photoionization of the negative charge state of $C_N$ and its conversion to the neutral charge state identified in prior work [10,11]. Since this feature was not distinguishable from the strong short wavelength absorption in other samples, the following discussion focuses on the three lower-energy peaks.

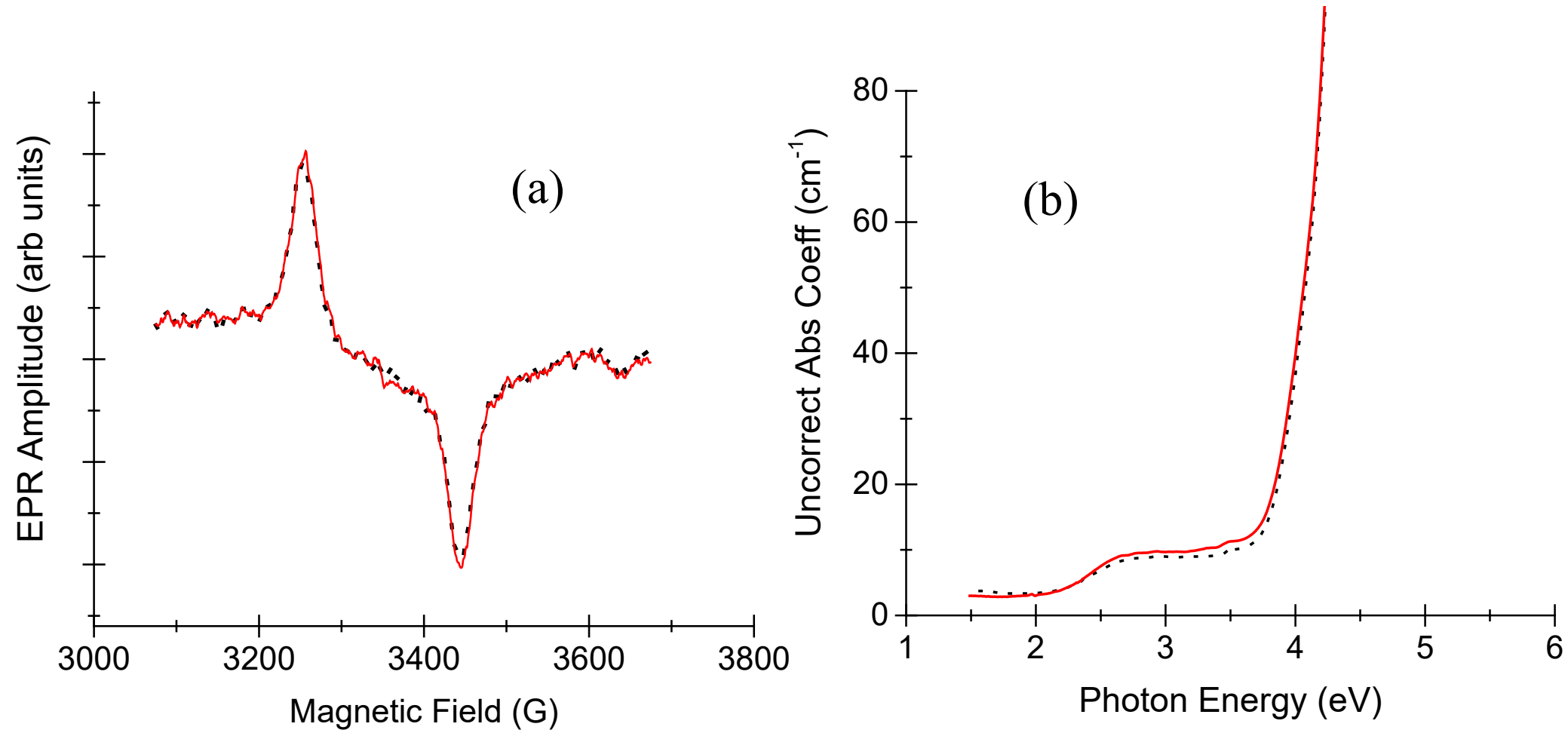

Figure 3. EPR (a) and optical absorption (b) spectra of an AlN sample from set 2: dark (black, dashed), after pre-illumination with 265 nm LED (red, solid).

EPR and optical absorption measurements were taken on samples in set 1 after illuminating with selected LEDs between 1200 nm and 265 nm (1 eV to 4.7 eV). Difference spectra were obtained by subtracting the dark measurement from each pre-illuminated spectrum, and amplitudes were normalized to the response from the 4.7 eV pre-illumination. The results are shown in Figure 2 as triangles (OA) and circles (EPR) where the different colors were obtained from different samples of set 1. The unfilled circles obtained from our earlier photo-EPR study are added here for clarity [9]. The correlation between the OA and EPR results is clear: the triangles (OA) and circles (EPR) follow the same trend over the entire band gap. Both show excitation above 4 eV and subsequent quenching between 1 and 3 eV. Furthermore, the amplitude of the OA and EPR signals decay at approximately the same rate - by about 40% percent over a period of 20 minutes at room temperature after removing the 4.7 eV pre-illumination. The comparable photo-induced behavior and post-illumination decay rates form the basis of a model discussed below.

The set 2 samples differ from those in set 1 in that $C_N^0$ is present in the dark and, as shown in Figure 3a, is minimally perturbed beyond the noise by exposure to the LED. The observation of $C_N^0$ in the dark indicates that the Fermi level is below the $C_N^{-/0}$ level, unlike set 1 where the Fermi level is above the $C_N^{-/0}$ level. This difference in Fermi level position that stabilizes different charge states of $C_N$ is likely due to a slight sample-to-sample variation in the relative concentration of donors with respect to the concentration of carbon. Figure 3b demonstrates that the OA between 2 eV and 4 eV is present in the dark and is only slightly altered by the 265 nm pre-illumination. The increase in OA is no more than 1 $cm^{-1}$, an order of magnitude less than that seen for set 1 samples. Although the absorption is broad, a fit to the data with three Gaussian line shapes led to peak energies within a few percent of those obtained for set 1. We also note that the slight change induced by the pre-illumination could be quenched with the 530 nm pre-illumination that returns the EPR signal to its initial state.

The photo-EPR and optical absorption spectroscopy performed on the same samples show that the OA between 2 eV and 4 eV is enhanced and quenched by pre-illuminating with the same wavelengths of light that are used to either quench or excite the EPR activity of $C_N$. Furthermore, when the $C_N$ EPR signal is minimally perturbed by the LEDs, the OA is also minimally perturbed under the same conditions. Taken together we arrive at the following conceptual view illustrated in Figure 4a of how the charge state of $C_N$ is related to the OA band that ranges from 2 eV to 4 eV.

Panels (1–4) describe sample set 1, where $C_N$ is initially in the negative charge state (EPR silent; panel 1). Pre-illumination at 4.7 eV photoionizes the occupied $C_N$ level (panel 2), converting it to the neutral charge state, quenching the 4.7 eV peak typically seen in OA spectra, and producing the EPR signature of the neutral charge state of $C_N$ (panel 3). The photoexcited electron is captured by a defect with transition level close to the conduction band before gradually recombining with $C_N$, accounting for the simultaneous slow decay of both the EPR signal and optical absorption seen experimentally. Before complete recombination, valence-band-to-$C_N$ transitions are optically accessible and the OA at 3.4 eV is observed (panel 4). Panels (5–6) describe the reverse process: pre-illumination at 2.3 eV returns $C_N$ to the negative charge state, quenching both the EPR signal and the 2–4 eV absorption.

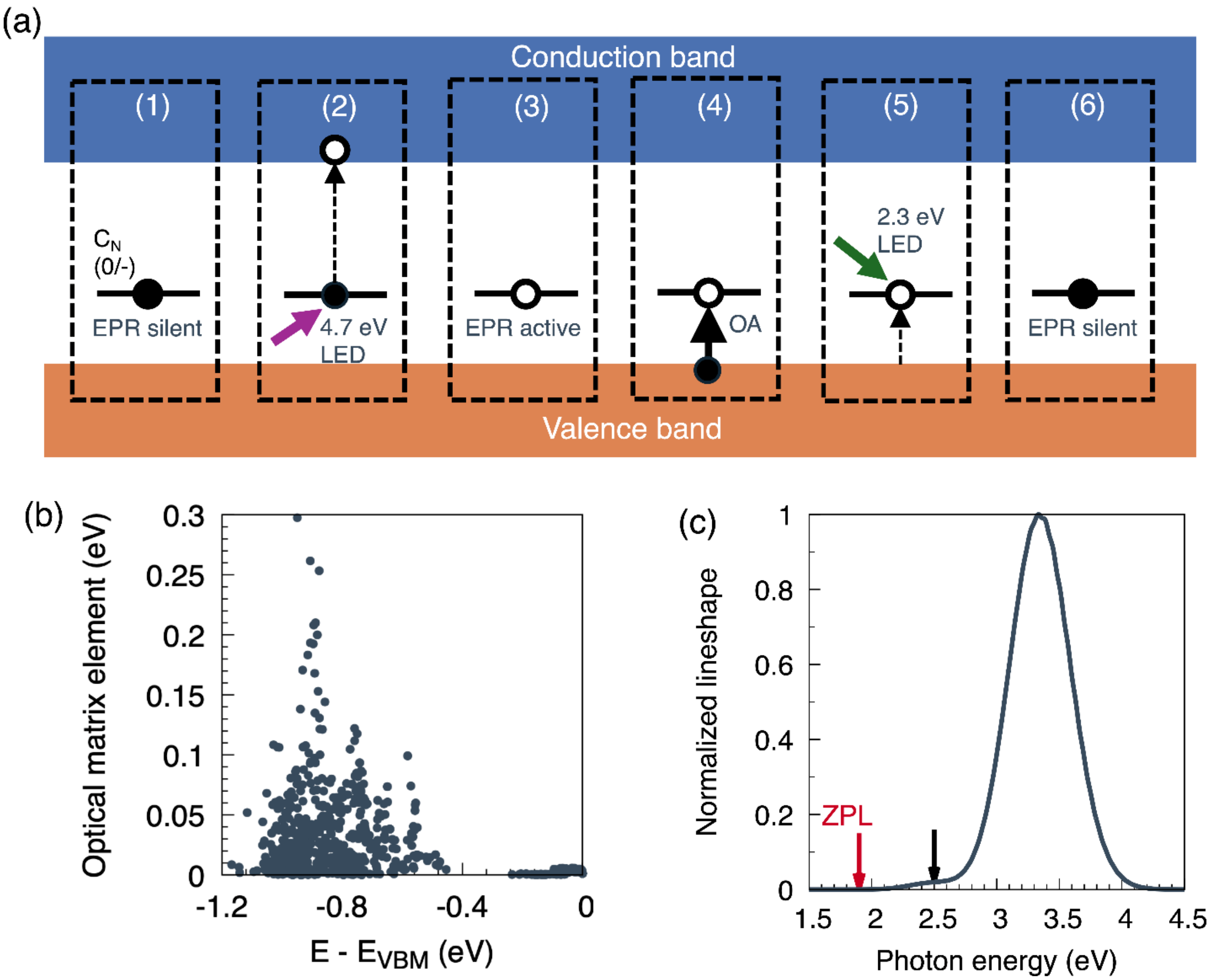

Figure 4. (a) Schematic illustration of the charge-state-dependent optical transitions of $C_N$ in AlN. Filled and unfilled circles denote occupied and unoccupied defect states, respectively. See main text for a description of the sequence of processes. (b) Calculated *k*-point-resolved optical matrix elements for optical transitions between valence band states and the $C_N$ acceptor level as a function of energy relative to the VBM. (c) Simulated optical absorption line shape accounting for both the Franck-Condon phonon broadening and the energy dependence of the oscillator strength shown in (b). The black vertical arrow indicates the peak position predicted if we assume the optical matrix elements are independent of energy (2.5 eV). The ZPL for the absorption process is illustrated with a vertical red arrow. The calculated line shape is normalized by the peak absorption value at 3.3 eV.

Within the Franck-Condon approximation, the peak absorption for the valence-band-to-$C_N$ transition is predicted at 2.5 eV [9], and one of the peaks we resolve in our fitting at 2.7 eV is consistent with this estimate. It is instructive to recall that the Franck-Condon approximation implicitly assumes the optical matrix elements are energy-independent across the valence band.

Figure 4b shows that this assumption breaks down. The optical matrix elements are strongly suppressed within ~0.5 eV of the VBM and peak approximately 1 eV below the VBM. This onset likely reflects the growing Al character of valence band states at deeper energies, enhancing spatial overlap with the $C_N$ defect wavefunction, which we have shown to hybridize strongly with the nearest-neighbor Al atom along the *c*-axis [9]. When this energy dependence is incorporated, we predict an absorption peak at 3.3 eV (Figure 4c). This is approximately 0.8 eV above the estimate if we were to assume the matrix elements are independent of energy, and in good agreement with the experimentally observed 3.4 eV peak.

Prior assignments of the 3.4 eV absorption to $V_{Al}$ and $V_{Al}$ –oxygen complexes have relied on first-principles calculations of transitions between these defects and the conduction band [14,16,17]. For $V_{Al}$ specifically, the relevant transition involves photoionization from the 3− charge state, which requires the Fermi level to lie ~3.1 eV above the VBM [14]. In our samples, however, the 3.4 eV absorption is observed only when $C_N$ is present in the neutral charge state, which is stable only when the Fermi level lies between 1.1 eV and 1.9 eV above the VBM [23]. For this range of Fermi energies, $V_{Al}$ and $V_{Al}$ –oxygen complexes are in the positive or neutral charge state, which allows us to rule out these defects as the origin of the 3.4 eV peak in our samples.

In summary, combining photo-EPR, optical absorption measurements, and first-principles calculations, we have identified the microscopic origin of a 3.4 eV absorption in AlN as a transition between the valence band and the $C_N$ acceptor level. This assignment is only possible when two factors are considered together: 1) the charge state of $C_N$, which, when directly identified by EPR determines whether the transition is optically accessible, and 2) the energy dependence of the optical matrix elements for transitions between $C_N$ and states below the valence band maximum using first-principles calculations. More broadly, this work emphasizes that reliable assignment of sub-bandgap absorption features in wide-bandgap semiconductors requires independent identification of the defect charge state combined with a theoretical treatment that captures the full energy dependence of the optical matrix elements.

This work was supported as part of the Ultra Materials for a Resilient Energy Grid, an Energy Frontier Research Center funded by the U.S. Department of Energy, Office of Science, Basic Energy Sciences under Award # DE-SC0021230. D.W. and J.L.L. were supported by the Office of Naval Research through the Naval Research Laboratory's Basic Research Program.

The authors have no conflicts to disclose.

The data that support the findings of this study are available from the corresponding author upon reasonable request.

1. Yijian Song, Rui He, Junxue Ran, Junxi Wang, Jinmin Li, Tongbo Wei, "III-nitride-based monolithic integration: From electronics to photonics", Appl. Phys. Rev. **12**, 021301 (2025).

2. Yu-Hsin Chen, Keisuke Shinohara, Jimy Encomendero, Naomi Pieczulewski, Kasey Hogan, James Grandusky, David A. Muller, Huili Grace Xing, Debdeep Jena, "From atomic-scale understanding to wafer-scale growth: Delta-doped AlN/GaN/AlN XHEMTs on single-crystal AlN by molecular beam epitaxy", APL Mater. **13**, 121103 (2025).

3. R. Sumathi, "Review—Status and Challenges in Hetero-epitaxial Growth Approach for Large Diameter AlN Single Crystalline Substrates" ECS Journal of Solid State Science and Technol. **10** 035001 (2021).

4. M. Kneissl, T. Seong, J. Han, H. Amano, "The emergence and prospects of deep-ultraviolet light-emitting diode technologies", Nature Photonics **13**, 233 (2019).

5. Yongzhou Xue, Hui Wang, Nan Xie, Qian Yang, Fujun Xu, Bo Shen, Jun-jie Shi, Desheng Jiang, Xiuming Dou, Tongjun Yu, and Bao-quan Sun, "Single-Photon Emission from Point Defects in Aluminum Nitride Films", J. Phys. Chem. Lett. **11**, 2689 (2020).

6. G.A. Cox, D.O. Cummins, K. Kawabe, R.H. Tredgold, "On the preparation, optical properties and electrical behaviour of aluminium nitride", J. Phys Chem. Solids **28**, 543 (167).

7. G. Slack and T Mcnelly, "AlN Single Crystals" Journal of Crystal Growth **42** 560 (1977).

8. M. Bickermann, B. Epelbaum, O. Filip, P. Heimann, S. Nagata, A. Winnacker, "Point defect content and optical transitions in bulk aluminum nitride crystals", Phys. Status Solidi B **246**, 1181 (2009).

9. Darshana Wickramaratne, Mackenzie Siford, Md Shafiqul Islam Mollik, John L. Lyons, and M. E. Zvanut, “Direct evidence for carbon incorporation on the nitrogen site in AlN”, Phys. Rev. Materials **8**, 094602 (2024).

10. Ramo´n Collazo, Jinqiao Xie, Benjamin E. Gaddy, Zachary Bryan, Ronny Kirste,Marc Hoffmann, Rafael Dalmau, Baxter Moody, Yoshinao Kumagai, Toru Nagashima, Yuki Kubota, Toru Kinoshita, Akinori Koukitu, Douglas L. Irving, and Zlatko Sitar, “On the origin of the 265 nm absorption band in AlN bulk crystals”, Appl. Phys. Lett. **100**, 191914 (2012).

11. Ivan Gamov, Carsten Hartmann, Thomas Straubinger, and Matthias Bickermann, “Photochromism and influence of point defect charge states on optical absorption in aluminum nitride (AlN)”, J. Appl. Phys. **129**, 113103 (2021).

12. Robert Rounds, Biplab Sarkar, Andrew Klump, Carsten Hartmann, Toru Nagashima, Ronny Kirste, Alexander Franke, Matthias Bickermann, Yoshinao Kumagai, Zlatko Sitar, and Ramón Collazo, “Thermal conductivity of single-crystalline AlN”, Applied Physics Express **11**, 071001 (2018).

13. A Sedhain, L. Du, J. Edgar, J. Lin, H. Jiang, “The origin of 2.78 eV emission and yellow coloration in bulk AlN substrates”, Appl. Phys. Lett. **95**, 262104 (2009).

14. Qimin Yan, Anderson Janotti, Matthias Scheffler, and Chris G. Van de Walle, “Origins of optical absorption and emission lines in AlN”, Appl Phys. Lett. **105**, 111104 (2014).

15. J.-M. Maki, I. Makkonen, F. Tuomisto, A. Karjalainen, S. Suihkonen, J. Raisanen, T. Yu. Chemekova, and Yu. N. Makarov, “Identification of the $V_{Al}$-$O_N$ defect complex in AlN single crystals” Phys. Rev. B **84**, 081204(R) (2011).

16. Joshua S. Harris, Jonathon N. Baker, Benjamin E. Gaddy, Isaac Bryan, Zachary Bryan, Kelsey J. Mirrielees, Pramod Reddy, Ramon Collazo, Zlatko Sitar, and Douglas L. Irving, “On compensation in Si-doped AlN”, Appl. Phys. Lett. **112**, 152101 (2018).

17. Qin Zhou, Zhaofu Zhang, and Baikui Li, Hui Li, Sergii Golovynskyi, Xi Tang, Honglei Wu, Jiannong Wang, and Baikui Li, “Below bandgap photoluminescence of an AlN crystal: Co-existence of two different charging states of a defect center”, APL Mater. **8**, 081107 (2020).

18. Richard C. Powell, *Physics of solid state laser materials* (Springer-Verlag New York Berlin Heidelberg, 1998).

19. W. Kohn and L. J. Sham, “Self-Consistent Equations Including Exchange and Correlation Effects”, Phys. Rev. **140**, A1133 (1965).

20. P. E. Blochl, “Projector augmented-wave method”, Phys. Rev. B **50**, 17953 (1994).

21. G. Kresse and J. Furthmuller, “Efficient iterative schemes for ab initio total-energy calculations using a plane-wave basis set”, Phys. Rev. B **54**, 11169 (1996); G. Kresse and J. Hafner, “Ab initio molecular-dynamics simulation of the liquid-metal–amorphous-semiconductor transition in germanium”, Phys. Rev. B **49**, 14251 (1994).

22. J. Heyd, G. Scuseria, and M. Ernzerhof, “Hybrid functionals based on a screened Coulomb potential”, J. Chem. Phys. **118**, 8207 (2003); J. Heyd, G. E. Scuseria, and M. Ernzerhof, “Erratum: “Hybrid functionals based on a screened Coulomb potential”, J. Chem. Phys. **124**, 219906 (2006).

23. J. L. Lyons, A. Janotti, and C. G. Van de Walle, “Effects of carbon on the electrical and optical properties of InN, GaN, and AlN”, Phys. Rev. B **89**, 035204 (2014).